\documentclass[twocolumn,amsmath,amssymb,prl,superscriptaddress,longbibliography]{revtex4-1}

\usepackage{titlesec}
\usepackage[colorlinks,bookmarks=false,citecolor=blue,linkcolor=red,urlcolor=blue]{hyperref}
\usepackage{epsfig}
\usepackage{color}
\usepackage{gensymb}
\usepackage{array}
\usepackage{subfigure}
\usepackage{graphicx} 
\usepackage{dcolumn}  
\usepackage{array,multirow,graphicx}
\usepackage{amsmath} 
\newcommand{\angstrom}{\textup{\AA}}

\newcommand{\be}{\begin{equation}}
\newcommand{\ee}{\end{equation}}

\newcolumntype{P}[1]{>{\centering\arraybackslash}p{#1}}

\begin{document}

\title{Single-site magnetic anisotropy governed by inter-layer cation charge imbalance in triangular-lattice AYbX$_2$ 
}

\author{Ziba Zangeneh}
\affiliation{Leibniz-Institute for Solid State and Materials Research, IFW Dresden, Helmholtzstr.~20, 01069 Dresden, Germany}

\author{Stanislav Avdoshenko}
\affiliation{Leibniz-Institute for Solid State and Materials Research, IFW Dresden, Helmholtzstr.~20, 01069 Dresden, Germany}

\author{Jeroen van den Brink}
\affiliation{Leibniz-Institute for Solid State and Materials Research, IFW Dresden, Helmholtzstr.~20, 01069 Dresden, Germany}
\affiliation{Department of Physics, Technical University Dresden, Helmholtzstr.~10, 01069 Dresden, Germany}

\author{Liviu Hozoi}
\affiliation{Leibniz-Institute for Solid State and Materials Research, IFW Dresden, Helmholtzstr.~20, 01069 Dresden, Germany}

\begin{abstract}
The behavior in magnetic field of a paramagnetic center is characterized by its $g$ tensor. An anisotropic form of the latter implies different kind of response along 
different crystallographic directions. Here we shed light on the anisotropy of the $g$ tensor of Yb$^{3+}$ 4$f^{13}$ ions in NaYbS$_2$ and NaYbO$_2$, 
layered triangular-lattice materials suggested to host spin-liquid ground states. 
Using quantum chemical calculations we show that, even if the ligand-cage trigonal distortions are significant in these compounds, the crucial role in realizing strongly anisotropic, 
``noncubic'' $g$ factors is played by inter-layer cation charge imbalance effects. The latter refer to the asymmetry experienced by a given Yb center due to having higher ionic charges 
at adjacent metal sites within the magnetic $ab$ layer, i.e., 3+ nearest neighbors within the $ab$ plane versus 1+ species between the magnetic layers. 
According to our results, this should be a rather general feature of 4$f^{13}$ layered compounds: less inter-layer positive charge is associated with stronger in-plane magnetic response.

\end{abstract}

\date\today
\maketitle

\section{Introduction}
Certain features of the magnetic interactions in Mott-Hubbard insulators may be anticipated already on the basis of details of their crystal structure. 
For example, for weak spin-orbit coupling (SOC), edge-sharing ligand octahedra with metal-ligand-metal bond angles close to 90$\degree$  are usually associated with ferromagnetic (FM) Heisenberg exchange 
while corner-sharing ligand octahedra (i.e., metal-ligand-metal bond angles of 180$\degree$ or somewhat less than 180$\degree$ when distortions are present) give rise to antiferromagnetic (AF) interactions. 
This trend is materialized in two of the Goodenough-Kanamori-Anderson rules for magnetic interactions in transition-metal oxides and halides \,\cite{Goodenough1958,Anderson1959,Kanamori1957,Kanamori57,Kanamori1959}.

Another structural detail that can affect the valence energy levels and magnetic couplings is the presence of large differences between cation charges in ternary and quaternary compounds. 
It turns out that charge `asymmetry' within the cation sublattice leads for instance to surprisingly large transition-metal $t_{2g}$-$e_g$ gaps in certain double perovskites\,\cite{Xu2016}
and to an anomalous sequence of the $t_{2g}$ levels and anomalous $g$ factors in layered square-lattice materials\,\cite{Bogdanov2015}. 
The precise charge and position of cation species in-between the magnetically active layers also strongly affect the strength of both isotropic and anisotropic intersite exchange couplings \,\cite{Yadav2018, Yadav2019}. 
Further, the interplay between cation charge imbalance in the immediate surroundings and distortions of the ligand cage allows in principle tuning of the single-ion anisotropy\,\cite{Bogdanov2013}.

While such effects have been addressed so far in transition-metal compounds, much less is known in this regard in the context of correlated $f$-electron systems. 
The $f$ orbitals being much tighter, it is not a {\it priori} clear if the asymmetry of the cation charge distribution in the nearby surroundings significantly affects $f$-shell states as well. 
Here we address such physics in the case of the $4f^{13}$ delafossites NaYbS$_2$ and NaYbO$_2$, layered triangular-lattice materials suggested recently to host spin-liquid ground states\,\cite{Baenitz2018, Baenitz2019, Tsirlin2019,Liu2018,Bordelon19}. 
We employ to this end quantum chemical calculations and reveal significant effects of this type on the $f$-level electronic structure. 
In particular, we find that, even if the trigonal distortions of the ligand cages are sizable in these compounds, the crucial role in realizing strongly anisotropic, noncubic $g$ factors is played by inter-layer cation charge imbalance. 
The latter competes with the effect of trigonal compression of the ligand octahedra, is strong enough to substantially modify the $f$-level splittings, 
and this way gives rise to highly anisotropic $g$ factors. 
Consequently, less inter-layer positive charge is associated with stronger in-plane magnetic response, a result that should be a rather general feature in 4$f^{13}$ layered compounds, 
intensively investigated nowadays as possible platforms for spin-liquid phases.

\section{Material model, computational scheme}
The discovery of the triangular-lattice magnet YbMgGaO$_4$ as a promising spin-liquid candidate\,\cite{Shen2016,Li15,YLi2015,Li2016,Paddison17,YLi2016} has generated further interest in the search for Yb-based systems with similar properties. 
Recently the ternary Yb-based delafossite-type compounds with the formula AYbX$_{2}$ (A\,=\,Na  and X\,=\,S, O) have gained a lot of attention in this context\,\cite{Baenitz2019,Baenitz2018,Tsirlin2019,Liu2018,Bordelon19}. 
Trigonally distorted YbX$_6$ octahedra share edges in these materials to form a two-dimensional triangular magnetic lattice of Yb$^{3+}$ ions. 
Successive layers are stacked along the crystallographic $c$ axis and separated by non-magnetic Na$^+$ species (see Fig. 1(a)). 
To describe the $f$-level electronic structure of Yb$^{3+}$ ions in this type of crystalline environment, we performed {\it ab initio} quantum chemical embedded-cluster calculations.
For this purpose, we used a finite atomic fragment as shown in Fig. 1(b), which consists of one YbX$_6$ reference octahedron, 
the six nearest-neighbor (NN) Yb atoms, and the first Na-ion coordination shell.  
The remaining part of the crystalline lattice was modeled as a large array of point charges fitted to reproduce the crystal Madelung field in the cluster region. The point-group symmetry is here  $D_{3d}$.

\begin{figure}
 \begin{center}
  \includegraphics[width=0.50\textwidth]{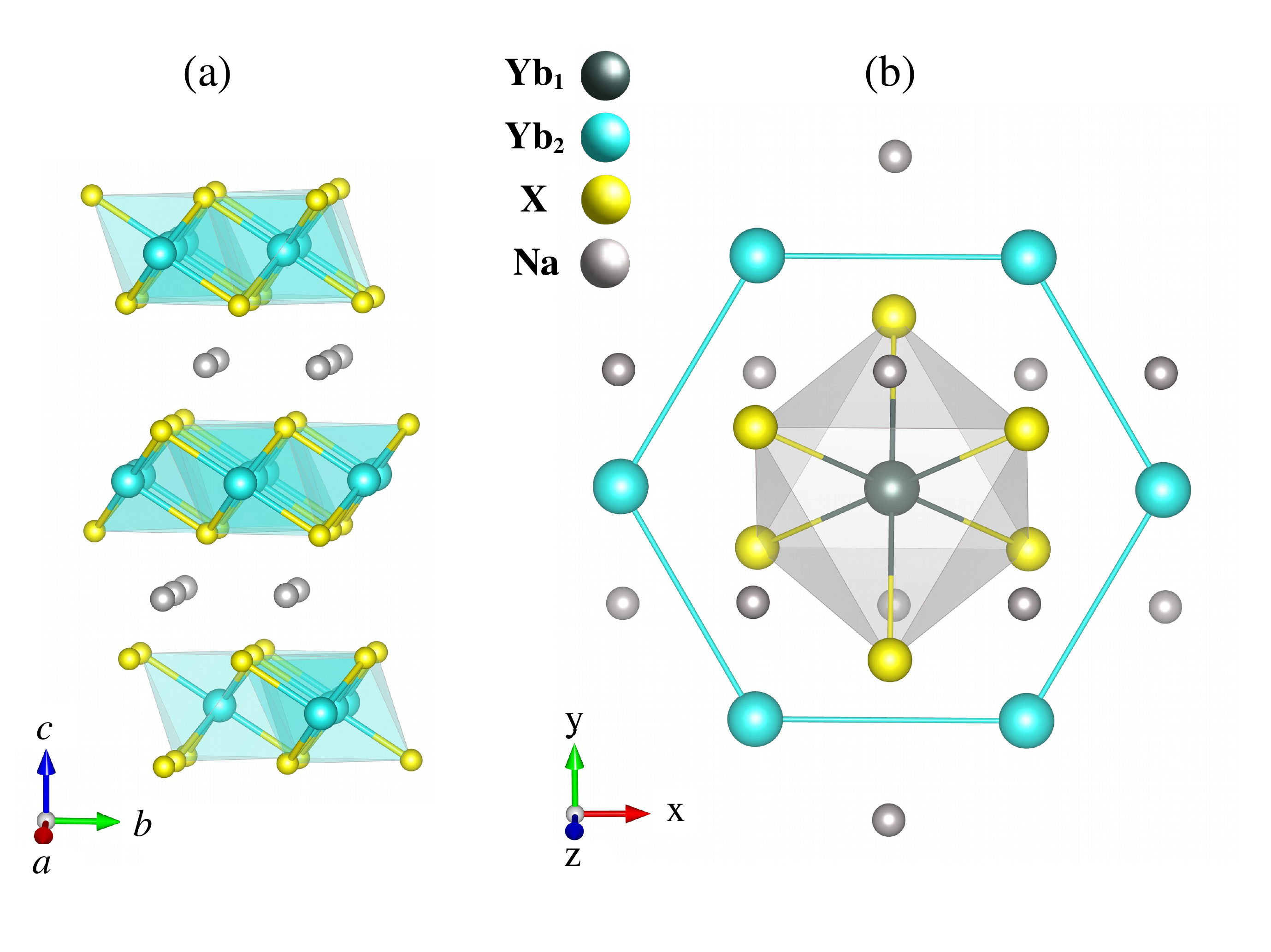}
  \caption{Crystal structure of NaYbX$_2$: (a) Stacking of YbX$_2$
triangular layers on top of each other. (b) The reference YbX$_6$ octahedron, the six NN Yb atoms, and the adjacent Na ions explicitly included in the calculations.
As our study is focused on single-site properties (Yb$_1$ central site), the adjacent six Yb ions (Yb$_2$) enter the calculations as large-core pseudopotentials\,\cite{ECPYb3} that include as well the $f$ electrons (see text for more details). 
The extended solid-state surroundings were modeled as an effective embedding potential.}
	\label{crystalStructure}
 \end{center}
\end{figure}

All quantum chemical calculations were performed with the {\sc molpro} package\,\cite{molpro12}.
For the initial complete-active-space self-consistent-field (CASSCF) optimization\,\cite{Helgaker2000}, we considered all seven $f$ orbitals at the central Yb site in the active space. 
Subsequently, the Yb $4f$ and S 3$s$, 3$p$  (or O 2$s$, 2$p$) electrons belonging to the central octahedron were correlated in multireference configuration interaction (MRCI)\,\cite{Helgaker2000} computations with single and double excitations. 
The reference CASSCF wavefunctions were optimized for all seven possible states associated with the 4$f^{13}$ manifold. 
All these eigenvectors entered the spin-orbit (SO) calculations, in both CASSCF and MRCI.
For the central Yb ion we employed energy-consistent relativistic pseudopotentials\,\cite{ECP} and Gaussian-type valence basis functions of quadruple-zeta quality \,\cite{ECPYb1,ECP1} while for the ligands
of the central YbX$_6$ octahedron we applied all-electron valence triple-zeta basis sets\,\cite{Woon1993,Dunning1989}.
The occupied shells of the Na ions were described as large-core pseudopotentials from the {\sc molpro} library\,\cite{ECPNa}.
To obtain a clear picture on crystal-field (CF) effects and SO interactions at the central Yb site, 
we cut off the magnetic couplings with the adjacent six Yb ions by applying large-core pseudopotentials that also incorporate the $4f$ electrons\,\cite{ECPYb3}. 

\section{Basic electronic structure}

Results for the $f$-level electronic structure are provided for NaYbS$_2$ and NaYbO$_2$ in Table I and Table II, using crystallographic data as reported in Refs.{\,\onlinecite{Schleid93}} and {\,\onlinecite{Hashimoto2003}}, respectively. 
Yb$^{3+}$ $4f^{13}$ CF splittings were first obtained by CASSCF and MRCI calculations without SOC.
Given the $D_{3d}$ point-group symmetry, the seven $f$ levels split into two groups of doubly degenerate $E_{u}$ and three non-degenerate $A_{2u}$ ($\times{2}$) and $A_{1u}$ ($\times{1}$) components \cite{Atkins70}. 
Without SO interactions we find the $^2A_{2u}$ state to be the lowest in energy for both NaYbS$_2$ and NaYbO$_2$ but our results indicate significantly larger CF splittings for the latter. 
This can be attributed to having shorter Yb-X distances (2.24 vs. 2.71{\,\angstrom}) and larger X-Yb-X angles (i.e., stronger trigonal compression, 96.3$\degree$ vs. 92.2$\degree$ X-Yb-X angles for ligands on the same side of the honeycomb plane) in NaYbO$_2$ as compared to NaYbS$_2$.

\begin{table}[b]
\caption{
CASSCF and MRCI results for the $f$-shell single-hole levels sans SOC in NaYbS$_2$ and NaYbO$_2$. Units of meV are used. 
In $D_{3d}$ point-group symmetry, the $f$ levels split into two groups of doubly degenerate $E_{u}$ and three non-degenerate $A_{2u}$ ($\times{2}$) and $A_{1u}$ ($\times{1}$) components. 
The order of the $^{2}A_{1u}$ and $^{2}E_{u}$ states is reversed in NaYbO$_2$ as compared to NaYbS$_2$.
}
\begin{tabular}{c|cc|cc}

\hline\\[-0.30cm]
\hline 
\multirow{2}{12em}{Yb$^{3+}$ $4f^{13}$ CF states} &\multicolumn{2}{c|}{NaYbS$_2$}         &\multicolumn{2}{c}{NaYbO$_2$ }  \\

          &\footnotesize{CASSCF}         &\footnotesize{MRCI}  &\footnotesize{CASSCF} &\footnotesize{MRCI}\\
\hline
\\[-0.10cm]
$   \!^{2}A_{2u}$ &0     &0     &0     &0    \\[0.30cm]
$   \!^{2}E_{u}$  &9     &10    &23    &27 \\ [0.30cm]  
$   \!^{2}A_{1u}$ &23    &29    &74    &85  \\[0.30cm]
$   \!^{2}E_{u}$  &31    &36    &68    &78 \\[0.30cm]
$   \!^{2}A_{2u}$ &40    &47    &106   &121\\[0.30cm]

\hline
\hline
\end{tabular}
\label{Na-CAS-CI-NoSOC}
\end{table}
Results of SO-CASSCF and SO-MRCI calculations are listed for NaYbS$_2$ and NaYbO$_2$ in Table II. 
The SO calculations were performed according to the methodology described in Ref.\,\onlinecite{SOC_molpro}, using Davidson-corrected\,\cite{Helgaker2000} MRCI energies as diagonal elements in the SO matrix.

Just as for a free $f^{13}$ ion, there is a large separation of more than 1 eV between the lowest four Kramers doublets ($^{2}F_{7/2}$ term in the case of the free ion) and the higher-lying states ($^{2}F_{5/2}$ term)\,\cite{Abragam1970}. 
In NaYbS$_2$, SO-MRCI puts the low-lying excited states at 15, 23, and 39 meV with respect to the ground-state Kramers doublet.
Higher excitation energies of 39, 46, and 92 meV  are obtained for NaYbO$_2$.

The SO-MRCI excitation energies are in reasonable agreement with available experimental data. For NaYbS$_2$, only two intense peaks originating from on-site $f$-$f$ transitions are observed in inelastic neutron scattering (INS), at 23 and 39 meV\,\cite{Baenitz2018}.
Additional information was obtained however by electron-spin resonance (ESR), i.e., a low-energy excitation at 17 meV\,\cite{Baenitz2018}. These experimental results are in very good correspondence with our data. For NaYbO$_2$, three intense peaks at 35, 58, and 83 eV are found by INS\,\cite{Tsirlin2019}, also in this case not far from our results. By ESR, the low-energy excitation comes at a smaller energy of about 27 meV\,\cite{Baenitz2019}. The details provided by our calculations will motivate further analysis for understanding the discrepancies between different experimental techniques.

It is worth to mention that earlier quantum chemical investigations of $4f$ systems were mainly carried out at the CASSCF/CASPT2 (complete active space second-order perturbation theory) level\,\cite{Seijo14,Pedersen14,Gendron15,Aiga14, Ning2012}. 
Here we report results of variational MRCI calculations: unlike non-variational methods, the calculated energy is in this case always higher than the actual energy. The calculation starts by choosing a `trial' wavefunction and any variations which lower its energy are necessarily making the approximate energy closer to the exact solution. The MRCI treatment shows significant corrections to the excitation energies (see Table II), percentagewise comparable to the corrections found in $d$-electron oxides\,\cite{Vamshi14,Huang11}.

\begin{table}[ht]
\caption{
Yb$^{3+}$ 4$f^{13}$ electronic structure as obtained by SO-CASSCF and SO-MRCI, in NaYbS$_2$ and NaYbO$_2$. Units of meV and notations for trigonal symmetry are used\footnote{See, e.g., Table 14 in Appendix B of Ref.\,\onlinecite{Abragam1970}.}. 
A gap of $\sim$1 eV similar to the free-ion $^{2}F_{7/2}$\,--\,$^{2}F_{5/2}$ splitting is clearly visible.
}
\begin{tabular}{p{1.1cm}p{0.8cm}|>{\centering\arraybackslash}p{1.6cm}>{\centering\arraybackslash}p{1.4cm}|>{\centering\arraybackslash}p{1.6cm}>{\centering\arraybackslash}p{1.4cm}}

\hline\\[-0.30cm]

\hline 
\multirow{2}{8em}{Yb$^{3+}$ $4f^{13}$ SO states}  & &\multicolumn{2}{c|}{NaYbS$_2$}         &\multicolumn{2}{c}{NaYbO$_2$} \\ 

 &    &\scriptsize{SO-CASSCF}  &\scriptsize{SO-MRCI}              &\scriptsize{SO-CASSCF}  &\scriptsize{SO-MRCI}\\
\hline
\\[-0.10cm]
$\ $        $\Gamma_{6}$            & &0      &0      &0       &0     \\[0.30cm]
$\ $        $\Gamma_{6}$            & &11     &15     &40      &39     \\ [0.20cm] 
$\ $        $\Gamma_{4}+\Gamma_{5}$ & &18     &23     &42      &46     \\[0.20cm]
$\ $        $\Gamma_{6}$            & &30     &39     &92      &92     \\[0.60cm]
$\ $        $\Gamma_{6} $           & &1277   &1300   &1320    &1319    \\[0.20cm]
$\ $        $\Gamma_{4}+\Gamma_{5}$ & &1282   &1305   &1324    &1326    \\[0.20cm]
$\ $        $\Gamma_{6} $           & &1300   &1329   &1383    &1383    \\[0.60cm]

\hline
\hline
\end{tabular}
\label{Na-CI-SOC}
\end{table}

\section{Cation charge imbalance effects}
Overall, our calculations (see Table I) indicate a sizable effect of trigonal distortions on the 4$f$ multiplet structure, in both NaYbX$_2$ compounds. 
While for two-dimensional Yb$^{3+}$ triangular-lattice systems such as NaYbS$_2$, NaYbO$_2$, and YbMgGaO$_4$, in the magnetic layer of Yb(O/S)$_6$ octahedra, the anion site is limited to a $2-$ ionic charge, in the non-magnetic layer(s) different substitution is possible\,\cite{Baenitz2019,Baenitz2018,Tsirlin2019,Li15,Shen2016}.
In order to gain better insight into the effect of the nearby surroundings on the 4$f$-level splittings, we performed additional calculations for hypothetical MgYbX$_2$ systems.
To this end, we replaced the inter-layer Na$^{1+}$ species by Mg$^{2+}$ cations, in both delafossite structures, NaYbS$_2$ and NaYbO$_2$\,\cite{Schleid93,Hashimoto2003}. 
Embedding potentials for fictitious lattice configurations with 2+ cations at all metal sites (Yb- and A-ion positions) were constructed and CASSCF and MRCI computations similar to those already discussed above were performed.
As we aim to maintain the reference-ion charge-state as Yb$^{3+}$  within a lattice of Mg$^{2+}$, Yb$^{2+}$, and X$^{2-}$ species, in order to keep overall charge neutrality, we added one negative charge within the nearby crystalline surroundings. 
Large-core pseudopotentials incorporating all occupied shells of the Mg ion were used\,\cite{ECPMg}. 
For the six Yb$^{2+}$ NN's, we again applied large-core pseudopotentials that also incorporate the $f$ shell\,\cite{ECPYb3}. 

Results for the Yb$^{3+}$ $f$-level splittings as obtained by CASSCF and MRCI sans SOC are provided for the hypothetical MgYbS$_2$ and MgYbO$_2$ systems in Table III. 
In comparison with the `real' Na-based compounds, the more homogeneous electrostatic surroundings in MgYbX$_2$ yield different CF splittings.
In particular, the latter are modified in MgYbS$_2$ such that the $^{2}A_{2u}$\,--\,$^{2}E_{u}$ and $^{2}A_{1u}$\,--\,$^{2}E_{u}$ gaps are reduced to only $\approx$2 meV.
This structure resembles now the $4f$ CF splittings in cubic symmetry, where the 4$f$ levels are grouped into two sets of triply-degenerate $T$ states and one non-degenerate $A$ component\,\cite{Li2008}. 
It looks as the effect of trigonal compression of the ligand cage is here cancelled out by the effect of the anisotropic (i.e., layered) crystalline surroundings. 
In MgYbO$_2$, on the other hand, a close-to-cubic CF level diagram is not achieved. 
But this is not at all surprising, given the different bond-lengths (metal-X and metal-metal) and the different amount of trigonal compression (X-Yb-X angles) in the S-based and O-based systems.

SO-CASSCF and SO-MRCI results for MgYbX$_2$ are shown in Table IV. 
As in the Na-based compounds, the $\Gamma_{6}$ state remains the lowest SO state for these more symmetric lattices with 2+ cations species in place. 
The MRCI treatment brings again sizable corrections to the CASSCF relative energies.
As concerns the MRCI data, the lower four Kramers doublets are at 0, 15, 17, and 33 meV in MgYbS$_2$. 
The $\Gamma_{4}+\Gamma_{5}$ state is split away from the $\Gamma_{6}$ state by only 2 meV, pointing again to a close-to-cubic 4$f$-shell electronic structure since in cubic symmetry the free-ion $J$\,=\,$7/2$ multiplet transforms into two $\Gamma_{6}$, $\Gamma_{7}$ doublets and one $\Gamma_{8}$ quartet\,\cite{Abragam1970}. 
In MgYbO$_2$, the low-lying four Kramers doublets are obtained at relative energies of 0, 24, 64, and 100 meV, with significantly larger splittings. 

\begin{table}[t]
\caption{
CASSCF and MRCI relative energies (meV) for the $f$-shell single-hole levels sans SOC for a hypothetical MgYbX$_2$ delafossite lattice of Mg$^{2+}$ (replacing Na$^{1+}$), Yb$^{2+}$, and X$^{2-}$ ions, 
with a more homogenous cation charge distribution.  
}
\begin{tabular}{c|cc|ccc}

\hline\\[-0.30cm] 
\hline 

Yb$^{3+}$ $4f^{13}$ states &\multicolumn{2}{c|}{MgYbS$_2$}         &\multicolumn{2}{c}{MgYbO$_2$ }  \\

 sans SOC               &\footnotesize{CASSCF}         &\footnotesize{MRCI}  &\footnotesize{CASSCF} &\footnotesize{MRCI}\\
\hline
\\[-0.10cm]
$   \!^{2}A_{2u}$ &0     &0     &0      &0    \\[0.30cm]
$   \!^{2}E_{u}$  &2     &2     &36     &37 \\ [0.30cm]  
$   \!^{2}A_{1u}$ &20    &26    &43     &50 \\[0.30cm]
$   \!^{2}E_{u}$  &23    &28    &101    &106\\[0.30cm]
$   \!^{2}A_{2u}$ &32    &41    &126    &134\\[0.60cm]
\hline
\hline
\end{tabular}
\label{Mg-CAS-CI-NoSOC}
\end{table}

\begin{table}[b]
\caption{
Yb$^{3+}$ $f^{13}$ SO states for a more symmetric cation charge distribution as realized in MgYbX$_2$ (X\,=\,S, O), with 2+ charges at all surroundings cation sites. Units of meV are used. 
}

\begin{tabular}{p{1.1cm}p{0.8cm}|>{\centering\arraybackslash}p{1.6cm}>{\centering\arraybackslash}p{1.4cm}|>{\centering\arraybackslash}p{1.6cm}>{\centering\arraybackslash}p{1.4cm}}

\hline\\[-0.30cm]

\hline 
\multirow{2}{6em}{Yb$^{3+}$ $4f^{13}$ SO states} & &\multicolumn{2}{c|}{MgYbS$_2$}     &\multicolumn{2}{c}{MgYbO$_2$} \\ 

 &    &\scriptsize{SO-CASSCF}  &\scriptsize{SO-MRCI}              &\scriptsize{SO-CASSCF}  &\scriptsize{SO-MRCI}\\
\hline
\\[-0.10cm]
 $\ $        $\Gamma_{6}$            & &0        &0      &0       &0     \\[0.20cm]
 $\ $        $\Gamma_{6}$            & &11       &15     &22      &24     \\  [0.20cm] 
 $\ $        $\Gamma_{4}+\Gamma_{5}$ & &14       &17     &62      &64     \\[0.20cm]
 $\ $        $\Gamma_{6}$            & &26       &33     &95      &100     \\[0.60cm]
 $\ $        $\Gamma_{6}$            & &1279     &1304   &1279    &1302    \\[0.20cm]
 $\ $        $\Gamma_{4}+\Gamma_{5}$ & &1279     &1305   &1310    &1331    \\[0.20cm]
 $\ $        $\Gamma_{6} $           & &1297     &1328   &1361    &1386     \\[0.60cm]

\hline
\hline
\end{tabular}
\label{Mg-CI-SOC}
\end{table}

Using the obtained SO wavefunctions we have also calculated the magnetic $g$ tensors of all compounds, using the Gerloch-McMeeking formula\,\cite{Bolvin2006}. 
The detailed computational scheme is described in Ref.{\,\onlinecite{Bogdanov2015}}. 
Ground-state $g$ factors for all AYbX$_2$ (A\,=\,Na, Mg and X\,=\,S, O) systems are listed in Table V, alongside with available experimental data. 

The MRCI  $g$ factors are $g_{ab}\,=\,3.19$, $g_{c}\,=\,0.93$ in NaYbS$_2$ and $g_{ab}\,=\,3.31$, $g_{c}\,=\,0.87$ in NaYbO$_2$, where the crystallographic $c$ axis is perpendicular to the $(ab)$ Yb honeycomb plane. 
Also according to ESR data\,\cite{Baenitz2018}, the $g$ factors
are highly anisotropic in NaYbS$_2$, in good agreement with our calculated $g$ factors. On the other hand, in NaYbO$_2$, the $g$ factor along the $c$ axis is underestimated in our
calculations with respect to the ESR measurements\,\cite{Baenitz2019}. The exact reason of this discrepancy is still unclear; however, given the different estimations for the position of the first excited state by ESR, INS\,\cite{Baenitz2018, Baenitz2019, Tsirlin2019}, and SO-MRCI, it was foreseeable to arrive to different $g$ factors in the latter compound.

It is seen that the most isotropic $g$ factors are computed in the MgYbS$_2$  system, namely $g_{ab}\,=\,2.73$, $g_{c}\,=\,2.49$ by SO-MRCI, and in MgYbO$_2$, $g_{ab}\,=\,3.09$, $g_{c}\,=\,3.25$. 
This again indicates that the MgYbS$_2$ potential is rather close to the cubic octahedral limit with $g^{cub}_{ab}\,=g^{cub}_{c}\,=2.67$.
The more isotropic structure of the $g$ factors is dictated by the redistribution of positive charge around the magnetic center: the Yb center experiences a more homogeneous electrostatic environment with 2+ ionic charges at NN sites both in the inter-layer region and within the magnetic plane.  

\begin{table}[ht]
\caption{
Ground-state $g$ factors in AYbX$_2$ (A\,=\,Na, Mg and X\,=\,S, O) as obtained from SO-CASSCF and SO-MRCI. Results as found by ESR are also provided.
}

\begin{tabular}{l|ll|ll|ll}
\hline
\hline
System    & \multicolumn{2}{c|}{CASSCF}                                               & \multicolumn{2}{c|}{MRCI}                                                 & \multicolumn{2}{c}{Experiment}                                                 \\
           & \multicolumn{1}{c}{$g_{ab}$}   & \multicolumn{1}{c|}{$g_c$} & \multicolumn{1}{c}{$g_{ab}$} & \multicolumn{1}{c|}{$g_c$} & \multicolumn{1}{c}{$g_{ab}$} & \multicolumn{1}{c}{$g_c$} \\ \hline
NaYbS$_2$ & 3.21                   & 0.80                    & 3.19                   & 0.93                    & 3.19                   & 0.57                   \\[0.20cm]
MgYbS$_2$ & 2.73                   & 2.49                    & 2.73                   & 2.49                    &                      &                     \\[0.20cm]
NaYbO$_2$ & 3.82                   & 1.01                    & 3.31                   & 0.87                    & 3.28                   & 1.75                   \\[0.20cm]
MgYbO$_2$ & 2.85                   & 4.26                    & 3.09                   & 3.25                    &                      &                      \\[0.20cm]
\hline
\hline                  
\end{tabular}
\label{g-factor}
\end{table}

\section{Conclusions}
To sum up, on the basis of the quantum chemical calculations, we provide a detailed analysis of the Yb$^{3+}$ 4$f^{13}$ electronic structure in NaYbS$_2$ and NaYbO$_2$. Significantly larger CF splittings are found in the latter, since the Yb-O bonds are shorter and the amount of trigonal distortion larger. The magnetic $g$ factors extracted from these computations are highly anisotropic. In particular, we obtained larger $g$ factors within the magnetic $ab$ layer. It turns out that this single-site magnetic anisotropy is largely controlled by cation charge imbalance in the nearby surroundings. According to our data, with a more symmetric cation charge distribution around the reference octahedron, the $f$-level splittings of the paramagnetic center are modified to a quasi-cubic energy diagram and the $g$ factors become more isotropic in both NaYbX$_2$ systems. The fact that in the sulphide especially the electronic structure brings back to the high-symmetry case when using more homogeneous surrounding charges, even if the octahedron is distorted and the local symmetry is $D_{3d}$, indicates an interesting situation with near cancellation between the effect of trigonal compression of the ligand cage and the effect of having anisotropic (i.e., layered) longer-range surroundings.
Our results suggest that the interplay of ligand-cage distortions, longer-range structural anisotropy, and cation charge imbalance in the immediate neighborhood provides means of varying the single-site $g_c$, $g _{ab}$ factors over a rather broad range.
More involved computations are required to determine how these knobs can be used for tuning the intersite magnetic couplings.

\section{Acknowledgments}
We acknowledge M. Baenitz, J. Sichelschmidt, A. Tsirlin, D. Inosov, Ph. Schlender, and K. Ranjith for inspiring discussions and U. Nitzsche for technical support. Z. Z., S. A., and L. H. acknowledge financial support from the German Research Foundation (Deutsche Forschungsgemeinschaft, DFG). This work was also supported by SFB1143 project A5 and the W\"urzburg-Dresden Cluster of Excellence ct.qmat.

\section{Bibliography}
\bibliography{biblio_AYbX2}

\end{document}